\documentclass[aps,preprintnumbers,amsmath,amssymb,prd,superscriptaddress,longbibliography,twocolumn]{revtex4-2}
\usepackage{dcolumn}
\usepackage{color}
\usepackage{dblfloatfix}
\usepackage[caption=false]{subfig}
\usepackage{feynmp}
\usepackage{feynmp-auto}
\usepackage{systeme,mathtools}
\usepackage{stmaryrd}
\usepackage{easyReview}

\def\comment#1{}

\newcommand{\nc}{\newcommand}
\nc{\beq}{\begin{eqnarray}}
	\nc{\eeq}{\end{eqnarray}}
\nc{\scs}{\scriptstyle}
\nc{\setval}{\fmfset{wiggly_len}{3mm} \fmfset{arrow_len}{1.5mm}
	\fmfset{arrow_ang}{13} \fmfset{dash_len}{1.5mm}\fmfpen{0.125mm}
	\fmfset{dot_size}{2thick}}

\usepackage{bm,latexsym,mathrsfs,enumerate,color}
\usepackage[mathcal]{euscript}
\usepackage[breaklinks=true,unicode=true,urlcolor = blue,colorlinks = true,citecolor = blue,linkcolor = blue]{hyperref}
\usepackage{graphicx}
\usepackage{todonotes}
\usepackage{wrapfig}

\renewcommand{\vec}[1]{\bm{#1}}

\def\slashchar#1{\setbox0=\hbox{$#1$}           
	\dimen0=\wd0                                 
	\setbox1=\hbox{/} \dimen1=\wd1               
	\ifdim\dimen0>\dimen1                        
	\rlap{\hbox to \dimen0{\hfil/\hfil}}      
	#1                                        
	\else                                        
	\rlap{\hbox to \dimen1{\hfil$#1$\hfil}}   
	/                                         
	\fi}                                         %

\DeclareMathAlphabet\mathbfcal{OMS}{cmsy}{b}{n}

\def\nablab{{\mbox{\boldmath $\nabla$}}}

\begin{document}
	
	\title{Axion electrodynamics of Weyl superconductors with broken time-reversal symmetry}
	
	\author{Vira Shyta}
	\affiliation{Institute for Theoretical Solid State Physics, IFW Dresden, Helmholtzstr. 20, 01069 Dresden, Germany}
	
	\author{Jeroen van den Brink}
	\affiliation{Institute for Theoretical Solid State Physics, IFW Dresden, Helmholtzstr. 20, 01069 Dresden, Germany}
	\affiliation{Institute for Theoretical Physics and W\"urzburg-Dresden Cluster of Excellence ct.qmat, TU Dresden, 01069 Dresden, Germany}
	
	\author{Flavio S. Nogueira}
	\affiliation{Institute for Theoretical Solid State Physics, IFW Dresden, Helmholtzstr. 20, 01069 Dresden, Germany}
	
	\begin{abstract}
		The low-energy effective description of Weyl semimetals is defined by the axion electrodynamics, which captures the effects arising due to the presence of nodes of opposite chirality in the electronic structure. Here, we explore the magnetoelectric response of time-reversal breaking (TRB) Weyl superconductors  in the London regime. The influence of the axion contribution leads to an increase in the London penetration depth --- a behavior that can be anticipated by first considering the photon spectrum of a TRB Weyl semimetal.  Moreover, we find that both the Meissner state and the vortex phase feature an interplay between the electric and magnetic fields.  This leads to a nonvanishing electromagnetic angular momentum, which we calculate for a number of geometrical configurations. 
	\end{abstract}
	
	\maketitle
	
\section{Introduction}

In a broad sense, axion electrodynamics is any theory of electromagnetic interactions whose action  contains the following term involving the scalar product between the electric and magnetic fields,  
\begin{equation}
	\label{Eq:Axion-action}
	S_a=\frac{\alpha}{4\pi^2}\int dt\int d^3r a(t,\vec{r}) \vec{E}\cdot\vec{B},
\end{equation}
where $\alpha=e^2/(\hbar c)$ is the fine structure constant, and $a(t,\vec{r})$ is the so-called axion scalar field \cite{Wilczek_PhysRevLett.58.1799}. Such a term is  time-reversal invariant  provided $a\to -a$ under a time-reversal transformation. In the past several years axion electrodynamics gained prominence in condensed matter physics, first through the establishment of effective topological field theory models for  time-reversal invariant  three-dimensional topological insulators (TIs) \cite{Qi_PhysRevB.90.045130,Ryu_PhysRevB.85.045104}, and then in studies of the electromagnetic response of Weyl semimetals \cite{Burkov-review}. Much of these developments borrow earlier ideas from studies of the axial anomaly in particle physics \cite{Adler_PhysRev.177.2426,bell1969pcac,Fujikawa_PhysRevD.21.2848,NIELSEN1983389}. Particularly influential in the condensed matter community is the paper by Nielsen and Ninomiya \cite{NIELSEN1983389}, which considers Weyl fermions in a crystal. 

The inclusion of the axion action in the usual electrodynamics leads to a modified form of the Maxwell equations \cite{Wilczek_PhysRevLett.58.1799},
\begin{equation}
	\label{Eq:Gauss-0}
	\nablab\cdot\left(\epsilon\vec{E}+\frac{\alpha a}{\pi}\vec{B}\right)=4\pi\rho,
\end{equation}
\begin{equation}
	\label{Eq:Ampere-0}
	\nablab\times\left(\frac{\vec{B}}{\mu}-\frac{\alpha a}{\pi}\vec{E}\right)=4\pi\vec{j}+\partial_t\left(\epsilon\vec{E}+\frac{\alpha a}{\pi}\vec{B}\right),
\end{equation}
while the source-free equations remain unaffected. The latter are only modified if magnetic monopoles are allowed, in which case even for a uniform $a$ the Maxwell equations get modified. Otherwise in case of a uniform axion field $a$ only the boundary conditions are modified; a fitting example of such behavior is offered by systems of TIs, where $a$ changes across surfaces that separate TIs from trivial insulators \cite{Qi-Hughes-Zhang_PhysRevB.78.195424}.  

In most condensed matter physics studies $a(t,\vec{r})$ is not a dynamical axion field, one notable exception being the study of topological superconductivity in Ref. \cite{Qi-Witten-Zhang_PhysRevB.87.134519}, where the axion field is equal to a difference between the phases of two superconducting order parameters. However, in TIs the axion field is typically assumed to be constant and when either time-reversal or inversion symmetry is present, it takes the value $a=\pi$. This follows from the fact that in  TIs $a$ is given by the flux of the Berry curvature in momentum space associated to the band structure of the crystal \cite{Qi-Hughes-Zhang_PhysRevB.78.195424}. This immediately results in a periodic axion field which upon application of discrete spacetime symmetries implies that $a=\pi$. The modified boundary conditions then lead to interesting physical effects, including the modification of optical properties that affect, for example, the Faraday and Kerr rotation \cite{Qi-Hughes-Zhang_PhysRevB.78.195424,Karch_PhysRevLett.103.171601}, which allowed to experimentally verify that the axion electrodynamics indeed describes  electromagnetic properties of TIs \cite{Armitage,Molenkamp}. TIs feature another optical effect important in the study of  plasmonics in these materials \cite{Karch_PhysRevB.83.245432}, namely, the axion Mie scattering on a spherical TI \cite{Lakhtakia:16,Lubk}, which can be generalized to other geometries. 

Interestingly, the axion electrodynamics also affects the electrostatic properties of the system. For instance, a point charge in the presence of a TI surface induces a magnetic field and a nonzero angular momentum of the electromagnetic field \cite{Zhang-monopoles,Nogueira_PhysRevResearch.4.013074}. Furthermore, surfaces separating TIs and conventional type-II superconductors cause fractional charges bounded to vortices to be induced at the interfaces, inducing an ac Josephson effect featuring an unconventional quantization of Shapiro steps \cite{Nogueira_PhysRevLett.117.167002} and a non-vanishing fractional angular momentum for the vortices \cite{Nogueira_PhysRevLett.121.227001}. The latter is a manifestation of the so-called Witten effect \cite{WITTEN1979283}, which posits that magnetic monopoles in the presence of a uniform axion leads to a dyon (a dipole made of electric and magnetic poles) with fractional charge. In the context of this paper this result easily follows from the modified Gauss law of Eq. (\ref{Eq:Gauss-0}). Pretty much the same Witten effect follows in the presence of vortices and in the absence of magnetic monopoles \cite{Nogueira_PhysRevLett.117.167002,Nogueira_PhysRevD.94.085003}. 

Contrary to the behavior of TIs, in Weyl semimetals the axion field  $a(t,\vec{r})$  is  not uniform due to the presence of Weyl nodes in the bulk \cite{Felser-review}. Generally, the axion field depends on both the momentum and energy separation of the Weyl nodes \cite{Burkov-review}, here denoted by $\vec{b}$ and $b_0$, respectively. The energy scales relevant in this effective theory are close to the Fermi energy allowing us to consider the Weyl cones irrespective of the way they actually connect to each other at higher
energies. A standard chiral anomaly calculation then leads to $a(t,\vec{r})=\vec{b}\cdot\vec{r}-b_0t$ \cite{Grushin_PhysRevD.86.045001,Burkov-review,Gorbar-review}. This form of the axion field underlies the fact that in principle in Weyl semimetals both the time-reversal invariance and inversion symmetry can be broken. However, if only the time-reversal invariance holds, $\vec{b}$ vanishes, while in a time-reversal broken (TRB) Weyl semimetal with the inversion symmetry preserved,  $b_0$ vanishes and $\vec{b}$ is nonzero. 

 Many quantum transport phenomena in Weyl semimetals have their origin in the chiral anomaly that manifests in the axion action \cite{Burkov_PhysRevB.86.115133, Spivak2013, PhysRevB.87.235306,Parameswaran2014}. And so a host of peculiar properties was  predicted in  Weyl semimetals and observed in the quantum and thermal transport. The findings include negative magnetoresistance, anomalous Hall effect, anomalous Nernst effect, gravitational anomaly and novel optical conductivity \cite{Hasan2015, ong_experimental_2021,Mertens2023,Magneto-optical-Weyl, Spivak2013}. Theoretically, the axion electrodynamics of Weyl semimetals should also lead to modifications in Faraday and Kerr rotation experiments \cite{COTE2023107064}.

In this paper we will focus on instances of axion electrodynamics that are connected to superconductivity, especially in Weyl systems. In this regard, we will adopt a minimal effective field theory strategy where the main physical ingredients appear due to symmetry considerations. For example, an introduction of a superconducting order parameter is simply dictated by a local $U(1)$ gauge invariance. Following this premise, superconductivity within the framework of axion electrodynamics is implemented as an Abelian Higgs model with the axion term given in Eq.~(\ref{Eq:Axion-action}). 

The most important physical quantity in the electrodynamics of superconductors (SCs) is the gauge invariant current density. 
Within axion electrodynamics of SCs, the total current density has to account for the axion action yielding, 
\begin{eqnarray}
	\label{Eq:j-SC-axion}
	\vec{J}&=&-iq(\psi^*\nablab\psi-\psi\nablab\psi^*)-2q^2|\psi|^2\vec{A}
	\nonumber\\
	&+&\frac{\alpha}{4\pi^2}(\partial_ta\vec{B}+\nablab a\times\vec{E}),
\end{eqnarray} 
where $q=2e$ is the charge of the Cooper pair and $\psi$ is the superconducting order parameter. 

Let us emphasize the essential physical content of Eq.~(\ref{Eq:j-SC-axion}) by considering the simplest realization of $U(1)$ symmetry breaking, corresponding to the London regime. Specifically, in its simplest incarnation, the London regime follows from the assumption of a frozen order parameter amplitude, $|\psi|=\sqrt{\rho_s}={\rm const}$, which represents the bare superfluid density. In addition, we also assume that the phase $\theta$ of the order parameter is gauged away, $\vec{A}\to\vec{A}+q^{-1}\nablab\theta$ (i.e., we are ignoring the vortices, so that $\nablab\times\nablab\theta=0$). Therefore, the current simplifies to,
\begin{eqnarray}
	\label{Eq:j-SC-axion-London}
	\vec{J}=-2q^2\rho_s\vec{A}
	+\frac{\alpha}{4\pi^2}(\nablab a\times\vec{E}+\partial_ta\vec{B}),
\end{eqnarray} 
Notably, the London current now does not only feature the familiar term (proportional to the vector potential) that leads to the Meissner effect, but also contains terms involving the magnetic and electric fields. In the scope of our effective field theory of Weyl semimetals, it is clear that Eq. (\ref{Eq:j-SC-axion-London}) takes the form, 
\begin{eqnarray}
	\label{Eq:j-SC-Weyl-London}
	\vec{J}=-2q^2\rho_s\vec{A}
	+\frac{\alpha}{4\pi^2}(\vec{b}\times\vec{E}-b_0\vec{B}). 
\end{eqnarray} 
 In the above expression for the current one recognizes two interesting  terms: a contribution proportional to the electric field and associated to the anomalous Hall effect \cite{Nagaosa-AHE,Burkov_2015,Burkov-review}, and another one proportional to the magnetic field, corresponding to the so-called chiral magnetic effect \cite{KHARZEEV2014133, Burkov_2015, Armitage2018}. The former is a consequence of time-reversal symmetry breaking, while the latter arises from inversion symmetry breaking.

In a similar vein, we have recently explored the electromagnetic response of time-reversal invariant (TRI) Weyl SCs \cite{Shyta-TRI-Weyl-SCs}. 
We found that in such systems an external magnetic field induces an additional component of the magnetic induction, which leads to helical
field profiles. This property is dictated by the axion contribution to the current density in Eq.~\eqref{Eq:j-SC-Weyl-London} (with $\vec{b}=0$ since the time-invariance is preserved). Remarkably, both vortex and Meissner states of TRI Weyl SCs are defined by a critical ratio
between the axion coupling and the London penetration depth. For a large enough axion coupling the vortex state becomes unstable and breaks down, and so a transition to the Meissner state ensues. The latter also features two drastically different responses depending on the aforementioned ratio: the state may feature helical magnetic fields that are screened in the superconducting bulk, or the screening breaks down and one anticipates oscillating magnetic induction. We dubbed this Meissner state “chiral”. 
A crucial physical quantity that characterizes the behavior of TRI Weyl SCs is the magnetic helicity. Interestingly, we will show in this paper that a similar role for TRB Weyl SCs is played by the electromagnetic angular momentum of the system.

Having recalled the instances of systems governed by effective field theories containing axion electrodynamics and the curious properties that lie therein, we move on to present our findings. Complementary to our previous work on TRI Weyl SCs, here we will investigate electrodynamics of Weyl superconductors with broken time-reversal invariance. This can be achieved in doped Weyl semimetals,  allowing to have a vanishing $b_0$ along with a Fermi surface around which a superconducting gap emerges. First, we will consider the photon excitation spectrum of Weyl semimetals, in particular the TRB ones. We will demonstrate that there is a gapped branch in the spectrum. We will show that one physical consequence of this gap leads to a magnetic induction profile reminiscent of the one in the London theory of superconductivity. Despite the similarity with the Meissner effect, such behavior is, strictly speaking, not superconducting. It simply reflects the topological mass generated by the anomalous Hall current. This motivates us to study the electrodynamics of TRB Weyl SCs, where both topological and Meissner screening coexist. We then proceed by investigating the Meissner state in a TRB Weyl SC using a simple slab geometry, similarly to canonical examples of London screening. An important feature emerging out of these calculations is the generation of an electric field due to the axion contribution to the effective action. This results in a nonvanishing electromagnetic angular momentum, which is a  property naturally underlying TRB Weyl systems. This is also reflected in the vortex state and gives rise to an anomalous Hall current contribution to the superconducting current density around the vortex line. Due to the induced electric field, this vortex state also carries an angular momentum, unlike conventional SCs.

\section{Topologically gapped photon excitation spectrum of Weyl semimetals}
\label{Sec:spectrum}

This section aims to demonstrate that the  axion contribution to the electrodynamics of Weyl semimetals leads to the gap generation in the photon excitation spectrum. As our main interest lies in magnetoelectric properties of Weyl superconductors, it is important to account for the peculiar axion electrodynamics of Weyl semimetals. Starting out with general arguments applicable to both time-reversal broken and time-reversal invariant systems, we will then derive in detail the photon spectrum in a TRB Weyl semimetal. This result will be illuminating since it will allow us to anticipate the screening of the magnetic field inside the TRB Weyl semimetal --- a feature present even in the absence of a superconducting phase. 

For convenience, we consider the partition function for the axion electrodynamics of Weyl semimetals within the imaginary time formalism, 
\begin{equation}
	\label{Eq:Z}
	Z=\int\mathcal{D}A_\mu\det(-\nabla^2)\exp\left[-\int_{0}^{\beta}d\tau\int  d^3r\left(\mathcal{L}_W+\mathcal{L}_{\rm gf}\right)\right],
\end{equation}   
where $A_\mu$ has components $(A_0,\vec{A})=(\phi,\vec{A})$, the electromagnetic Lagrangian for a Weyl semimetal is given by the sum of Maxwell terms and the axion contribution,
\begin{eqnarray}
	\label{Eq:L}
	\mathcal{L}_{W}&=&\frac{1}{8\pi}\left[\epsilon(\partial_\tau\vec{A}-\nablab\phi)^2+\frac{1}{\mu}(\nablab\times\vec{A})^2\right.\nonumber\\
	&+&\left.i\frac{2\alpha a}{\pi}(\partial_\tau\vec{A}-\nablab\phi)\cdot(\nablab\times\vec{A})\right],
\end{eqnarray}
while $\mathcal{L}_{\rm gf}=(\nablab\cdot\vec{A})^2/(2\lambda)$ is the Coulomb gauge-fixing term. Since the theory is Abelian, the Faddeev-Popov determinant is trivial, leading to the appearance of the factor $\det(-\nabla^2)$ in the integration measure of Eq. (\ref{Eq:Z}). This choice of the gauge is motivated by our interest in considering the static field equations later on, as the Coulomb gauge does not involve time.

The axion term invites us to perform the Gaussian integral over $\phi$ to obtain an effective action containing only the vector potential. The calculation is straightforward and leads to the effective Lagrangian, 
\begin{eqnarray}
	\label{Eq:Leff-Weyl}
	&&\!\!\!\!\!\!\!\!\mathcal{L}_{\rm eff}=\frac{1}{8\pi}\left[\epsilon(\partial_\tau\vec{A})^2+\frac{1}{\mu}(\nablab\times\vec{A})^2+i\frac{2\alpha a}{\pi}\partial_\tau\vec{A}\cdot(\nablab\times\vec{A})
	\right.\nonumber\\
	&+&\left.
	\frac{\alpha^2}{4\pi^3}\int d^3r'\frac{[\nablab a(\tau,\vec{r})\cdot\vec{B}(\tau,\vec{r})][\nablab' a(\tau,\vec{r}')\cdot\vec{B}(\tau,\vec{r}')]}{|\vec{r}-\vec{r}'|}\right],\nonumber\\
\end{eqnarray} 
where the Coulomb gauge has already been enforced, thus suppressing a term proportional to $[\partial_\tau(\nablab\cdot\vec{A})]^2$ in  the above expression.

The effective Lagrangian is interesting in general, especially in the case of a fully dynamical axion field theory (i.e., one where the axion field has an intrinsic dynamics via derivative and self-interaction terms in the action). However, here we will focus on Weyl semimetals with broken time-reversal invariance. In this case the axion field is simply given by $a(\vec{r})=\vec{b}\cdot\vec{r}$. It follows then that the effective Lagrangian of Eq. (\ref{Eq:Leff-Weyl}) can be written in the form, 
\begin{eqnarray}
	\label{Eq:Leff-Weyl-1}
	\!\!\!\!\!\!\!\!\mathcal{L}_{\rm eff}&=&\frac{1}{8\pi}\left[\epsilon(\partial_\tau\vec{A})^2+\frac{1}{\mu}(\nablab\times\vec{A})^2+i\frac{\alpha}{\pi}\vec{b}\cdot(\partial_\tau\vec{A}\times\vec{A})
	\right.\nonumber\\
	&+&\left.
	\frac{\alpha^2}{4\pi^3}\int d^3r'\frac{[\vec{b}\cdot\vec{B}(\tau,\vec{r})][\vec{b}\cdot\vec{B}(\tau,\vec{r}')]}{|\vec{r}-\vec{r}'|}\right],
\end{eqnarray}
where we have performed partial integrations  with respect to both $\tau$ and $\vec{r}$ in the third term.  

One immediate consequence of the above result is that the axion term leads to gapped branches in the spectrum of the photon \cite{Jackiw_PhysRevD.41.1231,Ferreiros_PhysRevB.93.195154}. This property is not immediately obvious by simply looking at the effective Lagrangian of Eq.~(\ref{Eq:Leff-Weyl}). The first thing we can do in order to understand the origin of the gapped photon, is  to consider the action for the axion term of Eq. (\ref{Eq:Axion-action}) in a covariant form, 
\begin{equation}
	\label{Eq:Axion-cov}
	S_a=-\frac{\alpha}{32\pi^2}\int d^4x a\varepsilon_{\mu\nu\lambda\rho}F^{\mu\nu}F^{\lambda\rho},
\end{equation}
where $F_{\mu\nu}=\partial_\mu A_\nu-\partial_\nu A_\mu$. Upon partial integration this becomes, 
\begin{equation}
	\label{Eq:Axion-cov-1}
	S_a=-\frac{\alpha}{16\pi^2}\int d^4x \varepsilon_{\mu\nu\lambda\rho}\partial^\mu a A^\nu F^{\lambda\rho},
\end{equation}
where we have assumed that the boundary terms vanish. This is an instructive exercise, since one can observe now that if $\partial_\mu a$ is a constant vector, the action above is akin to a Chern-Simons action. Therefore, this leads to a topologically massive gauge theory when the Maxwell action is also present \cite{DESER2000409}. 

Let us now discuss the gap generation in a more concrete way. To do this, we go back to the effective Lagrangian (\ref{Eq:Leff-Weyl-1}) of a TRB Weyl semimetal and assume, without loss of generality, that $\vec{b}=b\hat{\vec{z}}$. In this case, we can write the effective action in momentum space as, 
\begin{eqnarray}
	&&S_W = \frac{1}{8\pi\beta}\sum_n \int \frac{d^3p}{(2\pi)^3}\left[\epsilon \omega_n^2\delta_{ij} +\frac{1}{\mu}\left(\delta_{ij} p^2-p_ip_j\right)
	\right.\nonumber\\
	&+&\left.\frac{\alpha\omega_n}{\pi}\varepsilon_{ijk}b_k+\frac{\alpha^2}{\pi^2 \epsilon p^2}(\varepsilon_{imn}b_m p_n)(\varepsilon_{jkl}b_k p_l)\right]A_i(p)A_j(-p),
	\nonumber\\
\end{eqnarray} 
which in a matrix form  is given by,
\begin{equation}
	S_W = \frac{1}{8\pi\beta}\sum_n \int \frac{d^3p}{(2\pi)^3}A_i(p)M_{ij}(p)A_j(-p),
\end{equation}
with
\begin{widetext}
	\begin{equation}
		\hat{M}=	\left(\begin{array}{ccc}
			\frac{{b}^2 p_y^2 \alpha^2}{\pi^2 \epsilon p^2}+\frac{{p_y}^2+{p_z}^2}{\mu}+\epsilon \omega^2 & -\frac{{b}^2 {p_xp_y} \alpha^2}{\pi^2 \epsilon p^2}-\frac{{p_xp_y}}{\mu}+\frac{{b} \alpha \omega}{\pi} & -\frac{{p_xp_z}}{\mu} \\
			-\frac{{b}^2 {p_xp_y} \alpha^2}{\pi^2 \epsilon p^2}-\frac{{p_xp_y}}{\mu}-\frac{{b} \alpha \omega}{\pi} & \frac{{b}^2 {p_x}^2 \alpha^2}{\pi^2 \epsilon p^2}+\frac{{p_x}^2+{p_z}^2}{\mu}+\epsilon \omega^2 & -\frac{{p_yp_z}}{\mu} \\
			-\frac{{p_xp_z}}{\mu} & -\frac{{p_yp_z}}{\mu} & \frac{{p_x}^2+{p_y}^2}{\mu}+\epsilon \omega^2
		\end{array}\right).
	\end{equation}
Inversion of the above matrix results in the photon propagator whose poles yield the energy spectrum, 
\begin{equation}
	E_{ \pm}^2(p)=\frac{2 \pi^2 \epsilon p^4+b^2 \alpha^2 \mu\left(p^2+p_{\parallel}^2\right) \pm \alpha b \sqrt{4 \pi^2 \epsilon \mu p_z^2p^4+b^2 \alpha^2 \mu^2\left(p^2+p_{\parallel}^2\right)^2}}{2 \pi^2 \epsilon^2 \mu p^2},
\end{equation}
\end{widetext}
where $p^2_\parallel=p_x^2+p_y^2$. We note that one of the branches, namely, $E_-$  is gapless, since it vanishes for $p=0$, while $E_+$ does not.

For $\vec{p}=\vec{p}_\parallel$, the spectrum has the following two branches,
\begin{eqnarray}
	&&\!\!\!\!\!\! E_+(p_x,p_y, p_z=0)= \sqrt{c_b^2p_\parallel^2 +\frac{2\alpha^2 b^2}{\pi^2 \epsilon^2}},\\
	&&\!\!\!\!\!\! E_-(p_x,p_y, p_z=0)= c_b |\vec{p}_\parallel|,
\end{eqnarray} 
while for $\vec{p}=p_z\hat{\vec{z}}$,
\begin{eqnarray}
	E_{\pm}^2(p_\parallel=0, p_z)=c_b^2p_z^2+\frac{b^2 \alpha^2}{2 \pi^2 \epsilon^2}\\\pm \frac{b \alpha}{2\pi^2 \epsilon^2\mu}\sqrt{4\pi^2 \mu\epsilon p_z^2+b^2\alpha^2\mu^2},\nonumber
\end{eqnarray}
where $c_b=1/\sqrt{\epsilon\mu}$ is the speed of light in the bulk Weyl medium. Hence, as anticipated above, we have obtained gapped branches for the photon spectrum induced by the axion term. The gap is due to a term linear in $b$. A similar result has been derived earlier in the context of so-called helicons in Weyl semimetals \cite{Helicons-Pellegrino}.  In the next section we will see that a similar contribution appears in the electromagnetic response of both a TRB Weyl semimetal as well as a TRB Weyl superconductor.

\section{Weyl superconductors}

\subsection{Axion Abelian Higgs model}

The minimal model of a Weyl superconductor considered in this paper follows as a special case of an Abelian Higgs model with a Lagrangian (from now on we use units where $\hbar=c=1$),
\begin{equation}
	\label{Eq:L-Higgs}
	\mathcal{L}=\mathcal{L}_W+|(\partial_\mu+iqA_\mu)\psi|^2-m^2|\psi|^2-\frac{u}{2}|\psi|^4.
\end{equation} 
The field equations are straightforwardly obtained as, 
\begin{equation}
	\label{Eq:Gauss}
	\nablab\cdot\vec{E}=4\pi j_0-\frac{\alpha}{\pi}\nablab a\cdot\vec{B},
\end{equation}
\begin{equation}
	\label{Eq:Ampere}
	\nablab\times\vec{B}=4\pi\vec{j}+\frac{\alpha}{\pi}(\partial_ta\vec{B}+\nablab a\times\vec{E}),
\end{equation}
\begin{equation}
	\label{Eq:GL-like-equation}
	(\partial_\mu+iqA_\mu)^2\psi=-2m^2\psi+2u|\psi|^2\psi,
\end{equation}
where, 
\begin{equation}
	\label{Eq:j0}
	j_0=iq(\psi^*\partial_t\psi-\psi\partial_t\psi^*)-2q^2|\psi|^2A_0,
\end{equation}
\begin{equation}
	\label{Eq:j}
	\vec{j}=-iq(\psi^*\nablab\psi-\psi\nablab\psi^*)-2q^2|\psi|^2\vec{A}.
\end{equation}
We recognize on the RHS of Eq. (\ref{Eq:Ampere}) the current $\vec{J}$ of Eq. (\ref{Eq:j-SC-axion}).  

The above field equations are similar to the Ginzburg-Landau equations, except that they are written in spacetime rather than only in space, and also include the coupling to the axion field $a$. It is important to note the physical significance of the additional terms in the modified Maxwell equations (\ref{Eq:Gauss}) and (\ref{Eq:Ampere}). The Gauss law of Eq. (\ref{Eq:Gauss}) features an additional charge density which is induced by the magnetic induction due to axion term in the Lagrangian. Such a term is typically responsible for the emergence of the Witten effect \cite{WITTEN1979283}, which originally arises in the case where magnetic monopoles are present, implying in this way a fractionalization effect of the electric charge.


\subsection{London electrodynamics of Weyl superconductors}

Since the Abelian Higgs model can be viewed as an effective theory of superconductors, we will now apply the general formulas of the previous subsection to the case of Weyl superconductors. As we have discussed earlier, for Weyl semimetals we have $a(t,\vec{r})=\vec{b}\cdot\vec{r}-b_0t$. Inserting the latter into the Lagrangian (\ref{Eq:L-Higgs}) gives us a minimal effective model for the Higgs electrodynamics of Weyl superconductors. In order to analyze how the electrodynamics of SCs is influenced by the chiral anomaly, we will focus here on the London limit, which corresponds to a uniform $|\psi|$. Thus, we write $\psi=\sqrt{\rho_s}e^{i\theta}/\sqrt{2}$, where $\rho_s$ is the superfluid density. The phase $\theta$ is nonuniform and will play an important role in the vortex state. We have recently studied the London electrodynamics of Weyl SCs with time-reversal invariance \cite{Shyta-TRI-Weyl-SCs}, which was demonstrated to exhibit a chiral Meissner state. The latter occurs as a direct consequence of the chiral magnetic effect. Here we will focus on the London regime where time-reversal symmetry is broken, while inversion symmetry is preserved, which entails $b_0=0$. Thus, we are led to consider the following Lagrangian for a TRB Weyl SC,
\begin{eqnarray}
	\label{Eq:L-London}
	\mathcal{L}&=&\frac{\epsilon}{8\pi}\vec{E}^2-\frac{1}{8\pi}\vec{B}^2+\frac{q^2}{16\pi^2}(\vec{b}\cdot\vec{r})\left(\vec{E}\cdot\vec{B}\right)\nonumber \\
	& +&\frac{\rho_s}{2}\left[\left(\partial_t \theta+q \phi\right)^2-(\nabla \theta-q \vec{A})^2\right].
\end{eqnarray}
In the static regime, the generalized London equations that follow from this Lagrangian read,
\begin{eqnarray}
	\label{eq:fieldeq1bdotr}
	&&\!\!\!\!\!\!\!\!- \nablab\cdot \left(\epsilon \nablab \phi\right)+M^2 \phi=-\frac{q^2}{2 \pi} \vec{b} \cdot \vec{B},\\
	&&\!\!\!\!\!\!\!\!-\nabla^2 \vec{B}+M^2 \vec{B}=\frac{M^2}{q}\nablab\times\nablab \theta +\frac{q^2 }{2 \pi} \nablab\times \left(\vec{b}\times \vec{E}\right),
	\label{eq:fieldeq2bdotr}
\end{eqnarray}
where we introduced $M^2=4 \pi \rho_s q^2=1/\lambda^2$, with $\lambda$ being the London penetration depth. These field equations form a basis to investigate the electromagnetic response of TRB Weyl SCs.

In the next two subsections we will discuss the Meissner state and a vortex solution in simple geometries. 

\subsection{London electrodynamics of the Meissner state in TRB Weyl superconductors}  
\label{sec:slab-sol}

In the Meissner state there are no vortices and, therefore,  $\nablab\times\nablab\theta$ vanishes.  In this case, the London equations  \eqref{eq:fieldeq1bdotr} and \eqref{eq:fieldeq2bdotr} describing how the magnetic field penetrates the sample take the form,
\begin{eqnarray}
	\label{eq:bzeq1}
	&-& \frac{\partial}{\partial x}\left(\epsilon\frac{\partial \phi}{\partial x}\right) +M^2 \phi=-g B_z, \\
	&-&\frac{\partial^2}{\partial x^2} B_z+M^2 B_z=- \frac{\partial}{\partial x}\left(g\frac{\partial \phi}{\partial x}\right),
	\label{eq:bzeq2}
\end{eqnarray}
where we introduced $g=\frac{q^2b}{2\pi}$ and, similarly to the calculations in Sec. \ref{Sec:spectrum}, we chose $\vec{b}=b\vec{\hat{z}}$. 

The relevant boundary conditions are given by $B_z(x=0)=B_z(x=L)=B_{\rm ap}$ and
\begin{eqnarray}
	\label{eq:boundarycon1}
	&& -\left.\partial_x B_z\right|_{+\eta}=\left(g+\frac{\epsilon M^2}{g}\right) E_x(\eta) \\
	&& \left.\partial_x B_z\right|_{L-\eta}=- \left(g+\frac{\epsilon M^2}{g}\right)E_x(L-\eta),
	\label{eq:boundarycon2}
\end{eqnarray}
where $\eta\to0+$. The latter two conditions follow  from integrating Eqs.~\eqref{eq:bzeq1}  and \eqref{eq:bzeq2} with respect to $x$ on a narrow interval containing one of the surfaces of the SC. A crucial observation used in this calculation is that outside of the slab $\epsilon=1$ and $\vec{b}=0$. Eventually, considering  integration over $[-\eta, +\eta]$ for the surface situated at $x=0$ one  arrives at the boundary condition in Eq.~\eqref{eq:boundarycon1}. In the same vein,   Eq.~\eqref{eq:boundarycon2} is obtained by performing the integral around the surface at $x=L$. 

\begin{figure}[h]
	\subfloat[]{\includegraphics[width=0.5\linewidth]{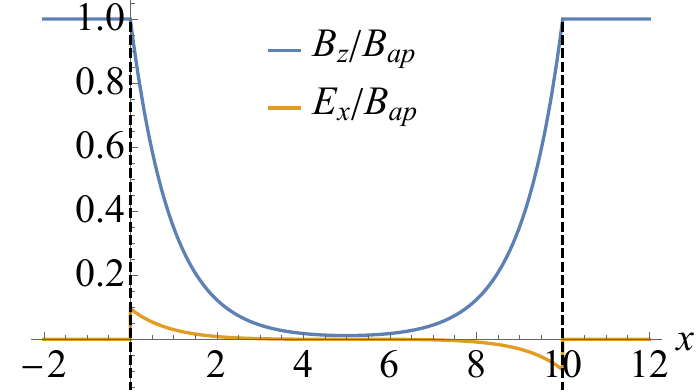}}
	\hfill
	\subfloat[]	{\includegraphics[width=0.5\linewidth]{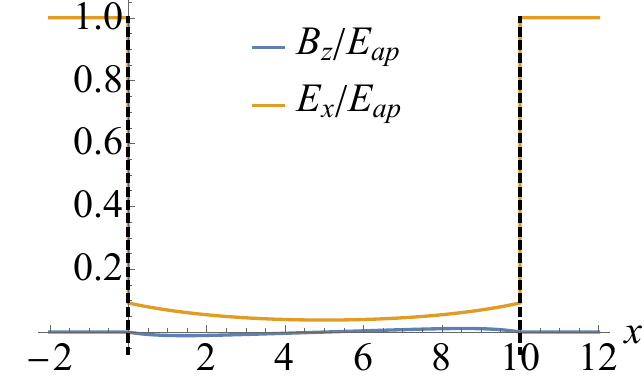}} 
	\hfill	
	\caption{Meissner state solutions in a TRB Weyl SC slab of finite thickness. One notices that due to the axion contribution the magnetic and electric fields are coupled.  Panel (a): the magnetic and electric field profiles inside the SC exposed to an external magnetic field $\vec{B}_\text{ap}=B_\text{ap}\hat{\vec{z}}$ [see Eqs.~\eqref{Eq:Bz-type-I-Bap} and \eqref{Eq:Ex-type-I-Bap}].  Panel (b): the magnetic and electric field profiles inside the SC exposed to an external electric field $\vec{E}_\text{ap}=E_\text{ap}\hat{\vec{x}}$ [see Eqs.~\eqref{eq:BzfromEap} and \eqref{eq:ExfromEap}].}
	\label{Fig:type-I}
\end{figure}

The solutions to Eqs.~\eqref{eq:bzeq1}  and \eqref{eq:bzeq2} are, 
\begin{equation}
	\label{Eq:Bz-type-I-Bap}
	B_z(x)=B_\mathrm{ap} \frac{\sum_{\sigma=\pm} \sigma c_\sigma \tau_\sigma  \sinh \left(\frac{\tau_\sigma L}{2}\right)\cosh \left(\tau_{-\sigma}\left(x-\frac{L}{2}\right)\right)}{\sum_{\sigma= \pm} \sigma c_\sigma \tau_\sigma \sinh \left(\frac{\tau_{\sigma } L}{2}\right) \cosh \left(\frac{\tau_{-\sigma} L}{2}\right)},
\end{equation}
\begin{equation}
	\label{Eq:Ex-type-I-Bap}
	E_x(x)=B_\mathrm{ap} g M^2 \sqrt{\epsilon} \frac{ \sum\limits_{\sigma= \pm} \sigma \tau_{-\sigma}^2 \sinh \left(\frac{\tau_\sigma L}{2}\right) \sinh \left(\tau_{-\sigma}\left(x-\frac{L}{2}\right)\right)}{\sum\limits_{\sigma= \pm} \sigma c_\sigma \tau_\sigma \sinh \left(\frac{\tau_\sigma L}{2}\right) \cosh \left(\frac{\tau_{-\sigma} L}{2}\right)},
\end{equation}
where for the sake of brevity we defined the constants,
\begin{widetext}
\begin{equation}
	\label{eq:taus}
	\tau_\pm^2=\frac{(1+\epsilon) M^2+g^2\pm \sqrt{\left(g^2+M^2(1+\epsilon)\right)^2-4 \epsilon M^4}}{2 \epsilon},
\end{equation}
\begin{equation}
	c_\pm=\frac{1}{2}\left[-g^4+M^4 \epsilon(1-\epsilon)-M^2 g^2(1+2 \epsilon)\pm\left(\epsilon M^2+g^2\right) \sqrt{\left(g^2+M^2(1+\epsilon)\right)^2-4 \epsilon M^4}\right].
	\label{eq:c-const-TRB}
\end{equation}

	We immediately see that the salient feature of this Meissner screening is the emergence of an electric field in the direction perpendicular to the surface of the slab. We also observe considerably more complex expressions for the field profiles, corresponding to a linear superposition of both magnetic inductions and electric fields  characterized by penetration lengths determined by $\tau_\pm$ given by Eq.  \eqref{eq:taus}. 
	The field profiles are depicted in Panel (a) of Fig.~\ref{Fig:type-I}.
	
	 The occurrence of the electric field in the solution above indicates that one could alternatively apply an external electric field, $\vec{E}_\mathrm{a p}=E_\mathrm{a p}\vec{\hat{x}}$, instead of $\vec{B}_\mathrm{a p}$. The resulting solution is illustrated by Panel (b) of Fig.~\ref{Fig:type-I} and explicitly given by
	\begin{equation}
	\label{eq:BzfromEap}
		B_z(x)=E_\mathrm{a p} g M^4 \frac{\sum\limits_{\sigma= \pm} \sinh \left(\frac{\tau_\sigma L}{2}\right) \sinh \left(\tau_{-\sigma}\left(x-\frac{L}{2}\right)\right)}{\sum\limits_{\sigma=\pm} \sigma c_\sigma \tau_\sigma \sinh \left(\frac{\tau_\sigma L}{2}\right) \cosh \left(\frac{\tau_{-\sigma}L}{2}\right)},
	\end{equation}
	\begin{equation}
	\label{eq:ExfromEap}
		E_x(x)=E_\mathrm{a p} M^2 \frac{\sum\limits_{\sigma= \pm}(-\sigma) \tau_\sigma\left(g^2+\epsilon M^2-\epsilon \tau_\sigma^2\right) \sinh \left(\frac{\tau_{-\sigma L}}{2}\right) \cosh \left(\tau_\sigma\left(x-\frac{L}{2}\right)\right)}{\sum\limits_{\sigma= \pm} \sigma c_\sigma \tau_\sigma \cosh \left(\frac{\tau_{\sigma }L}{2}\right) \sinh \left(\frac{\tau_{-\sigma } L}{2}\right)},
	\end{equation}
\end{widetext}
	where we used again the notation of Eqs.~\eqref{eq:taus} and~\eqref{eq:c-const-TRB}. It may seem from the effective axion action that in order to discuss effects stemming from axion electrodynamics one has to consider systems that feature components of $\vec{B}$ and $\vec{E}$ parallel to each other. However, the solutions above highlight that this is not necessarily required (i.e. an applied magnetic field induces an electric field perpendicular to it due to the axion contribution). This can also be observed directly from field equations  \eqref{eq:fieldeq1bdotr} and \eqref{eq:fieldeq2bdotr}.

In view of our results in Section \ref{Sec:spectrum} where a gapped branch of the photon was  shown to occur in a Weyl semimetal, it is instructive at this point to consider the special case where $M=0$. In this case Eqs. \eqref{eq:bzeq1} and \eqref{eq:bzeq2} can be combined into the equation, 
\begin{equation}
	-\frac{\partial^2B_z}{\partial x^2}+\frac{g^2}{\epsilon}B_z=0,
\end{equation}
which has the form of a conventional London equation.  However, this corresponds to the electromagnetic response of a TRB Weyl semimetal in a slab geometry, rather than a SC. Therefore, the standard London solution for the magnetic induction in a slab should follow,
\begin{equation}
	\label{Eq:Weyl-SM-B}
	\vec{B}(x)=B_{ \rm ap} \frac{\cosh \left(\frac{g}{\sqrt{\epsilon}}\left(x-\frac{L}{2}\right)\right)}{\cosh\left(\frac{g L}{2 \sqrt{\epsilon}}\right)}\hat{\vec{z}},
\end{equation}
as well the solution for the electric field, since the latter is induced, 
\begin{equation}
	\label{Eq:Weyl-SM-E}
	\vec{E}(x)=-B_{\rm a p}\frac{ \sinh \left(\frac{g}{\sqrt{\epsilon}}\left(x-\frac{L}{2}\right)\right)}{\sqrt{\epsilon} \cosh \left(\frac{g L}{2 \sqrt{\epsilon}}\right)}\hat{\vec{x}}.
\end{equation}
We also easily verify that these solutions correspond to the $M\to 0$ limit of the previously obtained ones for the Weyl SC, Eqs. \eqref{Eq:Bz-type-I-Bap} and \eqref{Eq:Ex-type-I-Bap}.  
This result has the following notable implication. Even in the absence of  superconductivity the magnetic field penetrating the TRB Weyl semimental is getting screened. On the other hand, if $\vec{B}_\text{ap}$ has components not parallel to $\vec{b}$, they do not resemble any screening behavior. This is indeed consistent with the results in Sec. \ref{Sec:spectrum}, where the photon spectrum of TRB Weyl semimetal was shown to exhibit gapped branches. This is a direct consequence of the topological mass following from the axion term in the action.  
These magnetoelectric properties are specific to TRB Weyl systems.

A further interesting property of this system is that in spite of being in a static regime, its electromagnetic field carries momentum (i.e., the Poynting vector is nonzero). As a consequence, also the electromagnetic angular momentum does not vanish. From Eqs.~\eqref{Eq:Weyl-SM-B} and \eqref{Eq:Weyl-SM-E} for a Weyl semimetal one obtains for a simple slab geometry the electromagnetic angular momentum per unit area. Calculating the integral $\vec{L}_\mathrm{EM} =(4\pi)^{-1} \int_{0}^{L}dx \vec{r}\times \left(\vec{E}\times \vec{B}\right)$ leads to,
\begin{equation}
	\label{Eq:Weyl-angular}
	\vec{L}_\mathrm{EM}=\frac{B^2_\mathrm{ap}L\sqrt{\epsilon}}{8\pi  g \coth\left(\frac{gL}{2\sqrt{\epsilon}}\right)} \left[ \coth \left(\frac{gL}{\sqrt{\epsilon}}\right)-\frac{\sqrt{\epsilon}}{gL}\right]\vec{\hat{z}}.
\end{equation}

\begin{widetext}
The TRB Weyl superconductor also possesses a nonvanishing electromagnetic angular momentum. This can already be seen from the slab solutions given in Eqs.~\eqref{Eq:Bz-type-I-Bap} and \eqref{Eq:Ex-type-I-Bap}. The corresponding angular momentum $\vec{L}_\mathrm{EM}$ per unit area of the slab will be directed along the $z$-axis and its component then can be calculated as 
\begin{equation}
	\begin{aligned}
		L_z^\mathrm{EM}	& =\frac{g M^2B_\text{ap}^2}{  \sum_{\sigma= \pm} 4\sigma c_\sigma \tau_\sigma \sinh \left(\frac{\tau_\sigma L}{2}\right) \cosh \left(\frac{\tau_{-\sigma} L}{2}\right)}\left\{\sum _ { \sigma = \pm } \left[-M^2 L c_\sigma \sinh ^2\left(\frac{\tau_\sigma L}{2}\right) \cosh \left(\tau_{-\sigma} L\right)+2 M^6 L \epsilon \sinh ^2\left(\frac{\tau_\sigma L}{2}\right)\right.\right. \\
		& \left.+\frac{\tau_\sigma  \epsilon^2 \sqrt{\epsilon}}{\left(g^2+M^2(1+\epsilon)\right)^2-4 \epsilon M^4}\left(\left(8 c_{-\sigma}+c_\sigma\right) \tau_{-\sigma}^4-6 c_\sigma \frac{M^4}{\epsilon}-3 c_\sigma \tau_\sigma^4\right) \sinh { }^2\left(\frac{\tau_\sigma L}{2}\right) \sinh \left(\tau_{-\sigma} L\right)\right] \\
		& \left.+M^6 L \sqrt{\epsilon}\left[\sinh \left(\tau_{+} L\right) \sinh \left(\tau_{-} L\right)-\sqrt{\epsilon} \cosh \left(\tau_{+} L\right) \cosh \left(\tau_{-} L\right)\right]\right\}, \\
	\end{aligned}
\end{equation}
where we used $\tau_\pm$ and $c_\pm$ defined in Eqs.~\eqref{eq:taus} and \eqref{eq:c-const-TRB}.  From this result it is clear that finite values of the angular momentum are only possible for $g\neq 0$. The crucial difference between the expression above and the angular momentum in Eq.~\eqref{Eq:Weyl-angular} lies in the presence of the superconducting order parameter captured by the inverse London penetration depth $M$, whose square is proportional to the superconducting stiffness $\rho_s$. The influence of superconductivity leads to a decrease in the value of $L_z^\text{EM}$ compared to the Weyl semimetal slab. This behavior is consistent with the fact that $M$ strengthens the screening of the electromagnetic fields in the slab.   
\end{widetext}

In order to discuss physical implications of this result, we turn to the spherical geometry.
 This way the superconducting sample is completely finite.  If one  considers a TRB Weyl SC sphere in an external magnetic field applied along the $z$-axis, an electromagnetic angular momentum is induced due to the axion contribution. This is interesting to compare to the  case of a spherical non-Weyl SC that is being rotated around the $z$-axis with the frequency $\vec{\omega}=\omega \vec{\hat{z}}$, where even in the absence of an external magnetic field a magnetic induction and electric field are generated inside the SC \cite{london1950superfluids, becker1933stromverteilung, Rotating-sphere-1964}. Hence, the rotation leads to a nonvanishing electromagnetic angular momentum.  This effect is also known as the London magnetic moment  occurring in a rotating non-Weyl superconductor \cite{london1950superfluids}. The London moment has direct applications and has been utilized in the Gravity Probe B relativity gyroscope experiment \cite{PhysRevLett.106.221101, Everitt_2015}.  

The behavior discussed above implies a certain equivalence between a   TRB Weyl SC in an external magnetic field and a rotating non-Weyl SC as there is an  interplay between contributions to the angular momentum from rotation and from magnetoelectric properties induced by the axion contribution. One can observe this already by considering a rotating TRB Weyl SC sphere and comparing it to a non-Weyl one. In the latter case, the rotation can be accounted for via an additional contribution to the current density in Eq.~\eqref{Eq:j}, which shifts the superfluid velocity by a term proportional to $\vec{\omega}\times \vec{r}$. Hence, on the right-hand side (RHS) of the standard field equation (given by Eq.~\eqref{eq:fieldeq2bdotr} for $\vec{b}=0$) a constant term linear in $\vec{\omega}$ appears. Then, using the boundary conditions that assume a dipole field configuration, one can solve the equation for $\vec{B}$. Already in a non-Weyl SC the external rotation plays a role of an applied magnetic field because it induces a nonvanishing magnetic induction directed parallel to  $\vec{\omega}$. While the $B$-field has an exponentially decaying profile, it does not vanish completely deep inside the SC, unlike the case where an external magnetic field is applied.  Instead,  the corresponding solution deep in the bulk of the non-Weyl  sphere is given by $\vec{B}=-2\omega\vec{\hat{z}}/q$ \cite{london1950superfluids, becker1933stromverteilung, Rotating-sphere-1964}.
	
 Following similar steps, one can consider a rotating TRB Weyl superconducting sphere. In this case, the same shift in the current density induced by rotation leads to the appearance of two additional constant terms on the RHS of Eq.~\eqref{eq:fieldeq2bdotr}. One of them is linear in $\vec{\omega}$ (and identical to the term discussed in the previous paragraph), while the other term arises from the influence of the rotation on the anomalous Hall contribution on the RHS of Eq.~\eqref{eq:fieldeq2bdotr}. The latter couples $\vec{\omega}$ and $\vec{b}$, and thus one observes traces of the presence of axion electrodynamics already considering the magnetic induction in the bulk of a rotating TRB Weyl  sphere, 	 $\vec{B}=\left[-2\omega/q+g\omega^2/(qM^2)\right]\vec{\hat{z}}$. An immediate implication following from this result is that there is a nonzero critical value of the rotation frequency that makes the magnetic induction inside the sample vanish, $\omega_c = 2M^2/g$. Therefore, the axion contribution allows to cancel out the effects of rotation on the field configuration, implying perfect diamagnetism. On the other hand, tuning the external rotation frequency in an experiment until the magnetic induction vanishes, could offer a way to measure the axion coupling.

\subsection{Vortex solutions of the axion London equation}  
\begin{figure}
	\centering
	\includegraphics[width=0.9\linewidth]{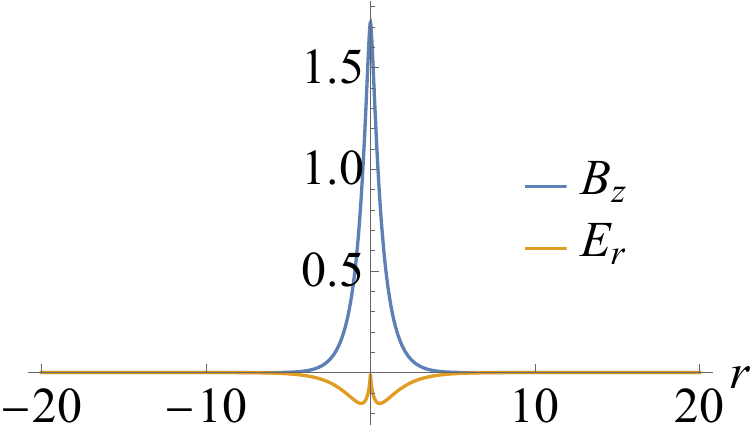}
	\hfill	
	\caption{Vortex state solutions in an infinite TRB Weyl SC. While the  magnetic induction given in Eq.~\eqref{eq:magneticfieldvortexbrinfinity1} has a behavior reminiscent of the usual result for a non-Weyl SC, a prominent feature of this solution is the rise of the electric field component perpendicular to the vortex line and given by Eq.~\eqref{Eq:E-infinite-vortex}. }
	\label{Fig:type-II-TRB}
\end{figure}

Having obtained and discussed the Meissner state solutions in a TRB Weyl SC, we proceed to derive vortex solutions in such SCs. Similarly to the previous section, we assume that $\vec{b}=b\hat{\vec{z}}$.
We  consider an infinite TRB Weyl SC with one vortex line located at the origin ($r=0$) along the $z$ direction. Therefore, $\nablab \theta$ has a singularity leading to nonvanishing vorticity $\nablab\times\nablab \theta = 2\pi n \delta^2(\vec{r})\hat{\vec{z}}$, where $n$ is the vortex winding number.  Accounting for this fact and using the Fourier transformation on Eq. \eqref{eq:fieldeq2bdotr}, we obtain an equation for the magnetic induction $B_z$,
\begin{equation}
	B_z(\vec{p})=\frac{2 \pi M^2 n\Phi_0 \delta\left(p_z\right)}{p_{\parallel }^2+M^2}+\frac{g p_{\parallel }^2 \phi(\vec{p})}{p^2+M^2},
\end{equation}
where $\Phi_0=2\pi/q$ is a magnetic flux quantum and $p^2 = p_\parallel ^2 + p_z^2$. Plugging this expression for the magnetic field into the second field equation \eqref{eq:fieldeq1bdotr}, we find the electric potential in the momentum space, 
\begin{equation}
	\phi(\vec{p})=-\frac{2 \pi n\Phi_0 g M^2 \delta\left(p_z\right)}{\epsilon\left( p_{\parallel}^2+\tau_+^2\right)\left(p_{\parallel}^2+\tau_{-}^2\right)}, 
\end{equation}
where $\tau_\pm$ were defined in Eq. \eqref{eq:taus}. Performing the inverse Fourier transformation, one obtains the electric field of the vortex line,
\begin{equation}
	\label{Eq:E-infinite-vortex}
	\vec{E}(\vec{r})=\frac{M^2 g n\Phi_0}{2 \pi \epsilon (\tau_+^2 -\tau_-^2)}\left[\tau_+ K_1\left(\tau_+ r\right)-\tau_{-} K_1\left(\tau_{-} r\right)\right] \vec{\hat{r}}
\end{equation}
as well as the magnetic induction,
\begin{widetext}
	\begin{equation}
		\label{eq:magneticfieldvortexbrinfinity1}
		\vec{B}(r)=\frac{n\Phi_0 M^2}{2 \pi} \left\{ K_0\left(Mr\right)+\frac{ g^2}{ \epsilon (\tau_+^2 -\tau_-^2)} \sum_{\sigma= \pm}\frac{\sigma \tau_{\sigma}^2}{\tau_\sigma^2-M^2}\left[K_0\left(M r\right)-K_0\left(\tau_{\sigma} r\right)\right] \right\}\vec{\hat{z}},
	\end{equation}
where  $K_\alpha(x)$ are modified Bessel functions of the second kind. From the magnetic induction expression we obtain the modified current density,
\begin{equation}
	\vec{j}(r) = \frac{n\Phi_0M^2}{8 \pi^2} \left\{ M K_1\left(Mr\right)+ \frac{ g^2}{ \epsilon (\tau_+^2 -\tau_-^2)}\sum_{\sigma= \pm}\frac{\sigma \tau_{\sigma}^2}{\tau_\sigma^2-M^2}\left[MK_1\left(M r\right)-\tau_\sigma K_1\left(\tau_{\sigma} r\right)\right]
	\right\}\vec{\hat{\varphi}}.
\end{equation}
The current density $\vec{j}$ gets an additional contribution due to the Weyl physics underlying this superconductor.
\end{widetext}
The obtained field profiles around the vortex line are depicted in Fig.~\ref{Fig:type-II-TRB}. Similarly to the Meissner state solution in Eqs.~\eqref{Eq:Bz-type-I-Bap} and~\eqref{Eq:Ex-type-I-Bap}, the magnetic induction expression is modified by the axion contribution and a radial electric field proportional to the axion coupling $g$ is induced.  The vortex solution can be adapted for the case of  a TRB Weyl SC slab of finite thickness. The corresponding results  given in Appendix~\ref{app:Weyl-slab-vortex} yield a similar picture, where electric and magnetic responses are coupled.

Similarly to the Meissner phase solutions discussed in the previous section, the vortex phase electromagnetic field profiles will also lead to a generation of a nonvanishing electromagnetic angular momentum, which will be directed along the vortex line.
In this case the electromagnetic angular momentum per length of the infinite vortex in a TRB Weyl SC is given by, 
\begin{widetext}
\begin{eqnarray}
	\label{Eq:L-vortex}
	L_z^\mathrm{EM}&=& \left(\frac{n \Phi_0}{2 \pi}\right)^2 \frac{gM^4}{\epsilon\left(\tau_{+}^2-\tau_{-}^{2}\right)}\left[\frac{2 \tau_{+}^2 }{\left(\tau_{+}^2-M^2\right)^2}\ln \left(\frac{\tau_{+}^2}{M^2}\right)-\frac{2 \tau_{-}^2 }{\left(\tau_{-}^2-M^2\right)^2}\ln \left(\frac{\tau_{-}^2}{M^2}\right)+\frac{M^2 }{\epsilon\left(\tau_{+}^2-\tau_{-}^2\right)^2}\ln \left(\frac{\tau_{-}^2}{\tau_{+}^2}\right)\right. \nonumber\\
	&-& \left.\frac{g^4+4 M^4(\epsilon-1)^2+5 g^2 M^2(1+\epsilon)+2 g^2 \epsilon\left(\tau_+^2-\tau_{-}^2\right)}{2 g^2 M^2\left(\tau_+^2-\tau_{-}^2\right)}\right],
\end{eqnarray} 
where the same definitions for $\tau_\pm$ given in Eq.~\eqref{eq:taus} were used. Once again we see that  $L_z^\mathrm{EM}$ is generated due to the axion contribution to the electromagnetic response. 
\end{widetext}

The prefactor $(n\Phi_0/2\pi)^2=n^2/q^2$ features a quantization of the angular momentum that is inherited from the flux quantization. A similar form of quantization occurs in the angular momentum of a vortex in the Chern-Simons Higgs electrodynamics \cite{Jackiw-Weinberg}. This is no accident. As we have commented earlier, the TRB form of the axion term in a Weyl semimetal does acquire a Chern-Simons form upon partial integration. This fact and the above expression for the angular momentum of the vortex highlights the topological nature of the problem.

\section{Conclusions and Outlook}

Here we considered the electromagnetic properties of Weyl superconductors with the broken time-reversal symmetry as described by the axion electrodynamics. What is interesting about this system is that the screening properties in absence of superconductivity exhibit a pattern quite similar to Meissner screening, despite the fact we are dealing with a semimetallic state. Hence, we have obtained superconducting solutions that reflect the interplay of two types of screening, one inherent to superconductivity itself (the Meissner effect) and another one following from the topological gap originating from the axion term when time-reversal symmetry is broken. The latter is akin to the type of screening occurring in Chern-Simons theories. 

One important physical conclusion following from our analysis is that any static solution of the field equations would upon applying an external magnetic field automatically generate besides a magnetic induction, also an electric field. We have studied here a few exemplary simple cases, e.g., a superconducting slab in an external magnetic field describing a Meissner and a single infinite vortex solution as well.    

Because of the characteristic pattern where both electric and magnetic fields are perpendicular to each other, a nonvanishing electromagnetic angular momentum is an intrinsic property in this type of system. This is somewhat reminiscent of earlier studies of Chern-Simons electrodynamics \cite{Jackiw-Weinberg}. In a similar vein, mechanical rotation of TRB Weyl systems, superconducting or not, can lead to effects transcending the realm of conventional rotating superconductors \cite{Becker, london1950superfluids, Hirsch2014}. Interestingly, we have shown that in view of the screening behavior of Weyl semimetals, there is a critical rotation frequency causing the magnetic induction to vanish in the bulk of the sample, thus implying perfect diamagnetism.     

\section*{ACKNOWLEDGMENTS}

We acknowledge financial support by the Deutsche Forschungsgemeinschaft (DFG, German Research Foundation), through SFB 1143 project A5 and the Würzburg-Dresden Cluster of Excellence on Complexity and Topology in Quantum Matter-ct.qmat (EXC 2147, Project Id No. 390858490). V.S. acknowledges the support by the German Academic Scholarship Foundation (Studienstiftung des deutschen Volkes).

\appendix
\begin{widetext}

\section{Vortex solutions for a slab of the time-reversal breaking Weyl superconductor of thickness $L$}
\label{app:Weyl-slab-vortex}
Here we consider the case of a TRB Weyl SC of finite thickness $L$, where a vortex line is perpendicular to the surface. The vortex solution for a slab defined in the region $|z|<L/2$ is more easily obtained by considering the differential equations for the vector potential.  Therefore, the London equation for the vector potential reads, 	
\begin{eqnarray}
	 A(r, z)&=&\frac{\Phi_0 M^2}{2 \pi} \left[ \left(1 - \frac{g^2 }{\epsilon \left(M_{+}^2-M^2\right)\left(M_{-}^2-M^2\right)}\right) \int_0 d p \frac{J_1(p r) a_1(p, z)}{p^2+M^2} \right. \\&-& \left.\frac{
		g^2}{\sqrt{\Delta}} \frac{1}{M_{+}^2-M^2} \int_0^\infty d p \frac{J_1(p r) a_2(p, z)}{p^2+M_+^2} + \frac{g^2}{\sqrt{\Delta}} \frac{1}{M_{-}^2-M^2} \int_0^\infty d p \frac{J_1(p r) a_3(p, z)}{p^2+M_-^2}\right] \nonumber
\end{eqnarray}
where  $J_\alpha(p r)$ are Bessel functions of the first kind and the function $a_j(p, z)$ is determined by application of the boundary conditions, which imposes the continuity of $\vec{A}$ and its derivatives with respect to $z$ at the interfaces $z=\pm L/2$. With this we obtain, 
\begin{equation}
	a_j(p, z)=\left\{\begin{array}{l}
		\frac{ \tau_j e^{p\left(\frac{L}{2}-z\right)}}{ \tau_j+p \operatorname{coth}\left(\frac{\tau_j L}{2}\right)}, \quad z>\frac{L}{2} \\
		1-\frac{p \cosh \left(\tau_j z\right)}{p \cosh \left(\frac{\tau_j L}{2}\right)+ \tau_j \sinh \left(\frac{\tau_j L}{2}\right)}, \quad-\frac{L}{2}<z<\frac{L}{2} \\
		\frac{ \tau_j e^{p\left(\frac{L}{2}+z\right)}}{ \tau_j+p \operatorname{coth}\left(\frac{\tau_j L}{2}\right)}, \quad z<-\frac{L}{2},
	\end{array}\right.
	\label{eq:a-func-TRB}
\end{equation}
with the notation, $\tau_j = \sqrt{p^2 + M^2_j}$, $j=+,-,0$. It is thus possible to derive the magnetic field profile of the vortex line in the slab of finite thickness, 
\begin{eqnarray}
	B_z&=& \frac{\Phi_0 M^2}{2 \pi} \left[ \left(1 - \frac{g^2 }{\epsilon \left(M_{+}^2-M^2\right)\left(M_{-}^2-M^2\right)}\right) \int_0 d p \frac{p J_0(p r) a_1(p, z)}{p^2+M^2} \right.\\&-& \left.\frac{
		g^2}{\sqrt{\Delta}} \frac{1}{M_{+}^2-M^2} \int_0^\infty d p \frac{pJ_0(p r) a_2(p, z)}{p^2+M_+^2} + \frac{g^2}{\sqrt{\Delta}} \frac{1}{M_{-}^2-M^2} \int_0^\infty d p \frac{p J_0(p r) a_3(p, z)}{p^2+M_-^2}\right] \nonumber
\end{eqnarray}
\begin{eqnarray}
	B_r&=& -\frac{\Phi_0 M^2}{2 \pi} \left[ \left(1 - \frac{g^2 }{\epsilon \left(M_{+}^2-M^2\right)\left(M_{-}^2-M^2\right)}\right) \int_0 d p \frac{J_1(p r) }{p^2+M^2} \frac{\partial a_1(p, z)}{\partial z} \right. \\&-& \left.\frac{
		g^2}{\sqrt{\Delta}} \frac{1}{M_{+}^2-M^2} \int_0^\infty d p \frac{J_1(p r)}{p^2+M_+^2}\frac{\partial a_2(p, z)}{\partial z}  + \frac{g^2}{\sqrt{\Delta}} \frac{1}{M_{-}^2-M^2} \int_0^\infty d p \frac{J_1(p r)}{p^2+M_-^2}\frac{\partial a_3(p, z)}{\partial z} \right],\nonumber
\end{eqnarray}
where 
\begin{equation}
	\frac{\partial a_j}{\partial z}=\left\{\begin{array}{l}
		\frac{ p\tau_j \operatorname{sgn}(z)e^{p\left(\frac{L}{2}-|z|\right)}}{ \tau_j+p \operatorname{coth}\left(\frac{\tau_j L}{2}\right)}, \quad |z|>\frac{L}{2} \\
		\frac{p \tau_j \sinh \left(\tau_j z\right)}{p \cosh \left(\frac{\tau_j L}{2}\right)+ \tau_j \sinh \left(\frac{\tau_j L}{2}\right)}, \quad |z|<\frac{L}{2},
	\end{array}\right.
\end{equation}
Next, since the axion contribution in a TRB Weyl SC couples magnetic and electric fields, we can calculate the electric potential,
\begin{equation}
	\phi(r, z)=\frac{g M^2}{\sqrt{\Delta}} \frac{\Phi_0}{2 \pi} \int_0^{\infty} d p p J_0(p r)\sum_{j= \pm}\frac{j\alpha_j(p, z)}{p^2+M_{j}^2},
\end{equation}
where the function $\alpha_j(p, z)$ defined by the boundary conditions differs from $a_j(p, z)$ in Eq.~\eqref{eq:a-func-TRB} by factors of $\epsilon$,
\begin{equation}
	\alpha_j(p, z)=\left\{\begin{array}{l}
		\frac{\epsilon \tau_j e^{p\left(\frac{L}{2}-z\right)}}{\epsilon \tau_j+p \operatorname{coth}\left(\frac{\tau_j L}{2}\right)}, z>\frac{L}{2}  \\
		1-\frac{p \cosh \left(\tau_j z\right)}{p \cosh \left(\frac{\tau_j L}{2}\right)+\epsilon \tau_j \sinh \left(\frac{\tau_j L}{2}\right)},-\frac{L}{2}<z<\frac{L}{2} \\
		\frac{\epsilon \tau_j e^{p\left(\frac{L}{2}+z\right)}}{\epsilon \tau_j+p \operatorname{coth}\left(\frac{\tau_j L}{2}\right)}, z<-\frac{L}{2}.
	\end{array}\right.
\end{equation}
Hence, the electric field components are given by 
\begin{equation}
	E_r=\frac{g M^2}{\sqrt{\Delta}} \frac{\Phi_0}{2 \pi} \int_0^{\infty} d p p^2 J_1(p r)\sum_{j= \pm}\frac{j \alpha_j(p, z)}{p^2+M_{j}^2},
\end{equation}
\begin{equation}
	E_z=\frac{g M^2}{\sqrt{\Delta}} \frac{\Phi_0}{2 \pi} \int_0^{\infty} d p p^2 J_0(p r)\left\{\begin{array}{l}
		\sum_{j= \pm} \frac{j\tau_j  \epsilon \operatorname{sgn}(z)}{p^2+M_j^2} \frac{e^{p\left(\frac{L}{2}-|z|\right)} }{\tau_j \epsilon+p \operatorname{coth}\left(\frac{\tau_jL}{2}\right)}, \quad \text{for $|z|>\frac{L}{2}$} \\
		\sum_{j= \pm} \frac{j\tau_j }{p^2+M_j^2} \frac{ \sinh \left(\tau_j z\right)}{p \cosh \left(\frac{\tau_jL}{2}\right)+\tau_j \epsilon \sinh \left(\frac{\tau_j L}{2}\right)}, \quad \text{for $|z|<\frac{L}{2}$}.
	\end{array}\right.
\end{equation}

\end{widetext}

\bibliography{weylSC}

\end{document}